\documentclass[structabstract]{aa}  
\usepackage{epsfig,graphicx,color}
\usepackage{graphicx}
\usepackage{natbib}
\usepackage{epstopdf}

\usepackage{txfonts}

\usepackage{txfonts}
\usepackage{natbib}
\begin{document}
\title{Water deuterium fractionation in the low-mass protostar NGC1333-IRAS2A
\thanks{Based on observations with the APEX telescope, the IRAM 30m telescope and with the James Clerk Maxwell Telescope (JCMT). APEX is a collaboration between the Max-Planck-Institut f\"ur Radioastronomie, the European Southern Observatory, and the Onsala Space Observatory. IRAM is a European collaboration between the CNRS (Centre National de la Recherche Scientifique), the MPG (Max-Planck-Gesellschaft) and the Spanish IGN (Instituto Geogr‡fico Nacional). 
The JCMT is operated by The Joint Astronomy Centre on behalf
of the Particle Physics and Astronomy Research Council of the United Kingdom,
the Netherlands Organization of Scientific Research, and the National Research
Council of Canada.}}

      \subtitle{}
      \titlerunning{water deuterium fractionation in NGC1333-IRAS2A}
      \author{F.-C. Liu\inst{1}\thanks{Member of the International Max Planck Research School (IMPRS) for Astronomy and Astrophysics at the Universities of Bonn and Cologne.}
   		 \and
   		 B. Parise\inst{1}
          \and
          L. Kristensen\inst{2}
          \and
          R. Visser\inst{2}
          \and
          E.F.~van~Dishoeck\inst{2,3}
          \and
          R. G\"{u}sten\inst{1}
                       }

      \institute{Max-Planck-Institut f\"ur Radioastronomie, Auf dem H\"ugel 69, 53121 Bonn, Germany\\
              \email{fliu@mpifr-bonn.mpg.de}
         \and Leiden Observatory, Leiden University, PO Box 9513, 2300 RA Leiden, The Netherlands
		\and Max-Planck-Institut f\"{u}r Extraterrestrische Physik, Giessenbachstrasse 1, 85748 Garching, Germany
}
      \date{Received Aug xx, 2010; accepted Oct yy, 2010}

        \abstract
{Although deuterium enrichment of water may provide an essential piece of information in the understanding of the formation of comets and protoplanetary systems, only a few studies up to now have aimed at deriving the HDO/H$_2$O ratio in low-mass star forming regions. Previous studies of the molecular deuteration toward the solar-type class 0 protostar, IRAS 16293-2422, have shown that the D/H ratio of water is significantly lower than other grain-surface-formed molecules. It is not clear if this property is general or particular to this source.}
 {In order to see if the results toward IRAS 16293$-$2422 are particular, we aimed at studying water deuterium fractionation in a second low-mass solar-type protostar, NGC1333-IRAS2A.}
 {Using the 1-D radiative transfer code RATRAN, we analyzed five HDO transitions observed with the IRAM 30m, JCMT, and APEX telescopes. We assumed that the abundance profile of HDO in the envelope is a step function, with two different values in the inner warm (T$>$100\,K) and outer cold (T$<$100\,K) regions of the protostellar envelope.}
 {The inner and outer abundance of HDO is found to be well constrained at the 3$\sigma$ level.
The obtained HDO inner and outer fractional abundances are x$^{\rm \tiny{HDO}}_{\rm in}=6.6\times10^{-8}$--$1.0\times10^{-7} (3\sigma)$ and x$^{\rm \tiny{HDO}}_{\rm out}=9\times10^{-11}$--$1.8\times10^{-9} (3\sigma)$. These values are close to those in IRAS 16293-2422, which suggests that HDO may be formed by the same mechanisms in these two solar-type protostars. Taking into account the (rather poorly constrained) H$_2$O abundance profile deduced from $Herschel$ observations, the derived HDO/H$_2$O in the inner envelope is $\geq$ 1\% and in the outer envelope it is 0.9\%--18\%. These values are more than one order of magnitude higher than what is measured in comets. If the same ratios apply to the protosolar nebula, this would imply that there is some efficient reprocessing of the material between the protostellar and cometary phases.}
 { The H$_2$O inner fractional abundance could be further constrained by an analysis of newer observations of high-energy H$_2^{18}$O lines. This new observations would be required to understand water fractionation in more detail.}


      \maketitle
%
\section{Introduction}
Molecular deuteration has been the focus of many studies. The abundance of deuterated molecules has been shown to be significantly increased in low-temperature environments, compared to the D/H elemental ratio in the interstellar medium \citep[$\sim$1.5$\times$10$^{-5}$,][]{Linsky03}. Over the last decade, very high D/H ratios in several molecules have been found in low-mass protostars, e.g. D$_2$CO/H$_2$CO=0.05 \citep{Ceccarelli98} and CD$_3$OH/CH$_3$OH=0.01 \citep{Parise04} toward IRAS 16293-2422. This can be interpreted in terms of the chemical formation pathways of these molecules -- both are believed to be grain surface products in cold dark clouds through the formation process of successive CO hydrogenation \citep{parise06a}.
Because water is the main constituent of grain mantles, studies have been performed to investigate if deuterium enrichment of water is similar to the fractionation of formaldehyde and methanol. Several HDO transitions have been observed toward IRAS 16293$-$2422, and were modeled to derive the HDO abundance profile in its envelope \citep{Parise05a}. Using the H$_2$O abundance profile as derived from ISO observations \citep{Ceccarelli00a}, the HDO/H$_2$O\, ratio was found to be $\sim$\,0.03 in the inner hot corino, and 
$<$0.002 in the cold outer envelope \citep{Parise05a}, i.e. significantly lower than the formaldehyde and methanol fractionation in the same source. A search for HDO ices in low-mass sources with large fractionation of formaldehyde and methanol found no HDO ices to a very low limit \citep[HDO/H$_2$O $<$ 0.01 in ices]{Parise03}, which might be surprising if all species formed simultaneously on dust surfaces. 
 
This low deuterium enrichment of water, if confirmed, would be a very valuable constraint for astrochemical models that strive to explain the chemical processes involved in the formation of water.  
Studying the abundance and distribution of HDO in low-mass prestellar envelopes is moreover an essential astrochemical goal, as it would provide the early conditions from where comets form. The HDO/H$_2$O ratio has been measured in several comets and found to be $\sim$\,3$\times$10$^{-4}$ \citep{Balsiger95, BockeleeMorvan98, Meier98a}. Up to now, the HDO emission has been studied in detail only in one low-mass protostar \citep[IRAS 16293$-$2422,][]{Parise05a}. A radiative transfer analysis of HDO emission has shown that the abundance profile of HDO can be constrained in low-mass protostellar envelopes by means of the analysis of several HDO lines, spanning different energy conditions \citep{Parise05b}. 
Now that the $Herschel$ telescope is delivering its first data, which allows us to get a better measure of the water abundance profile in protostars, the time is ripe to study water fractionation in these environments.
Here we present the modeling of HDO observations toward a second solar-type Class 0 protostar, NGC1333-IRAS2A (IRAS2A). In Section 2 we present the observations and first results, in Section 3 we present the radiative transfer modeling, and discuss the results in Section 4. We finally conclude in Section 5. \\


\section{Observations and results}
\subsection{The source} \label{source}
IRAS2A is a solar-type Class 0 protostar located in the NGC 1333 molecular cloud. The adopted distance of IRAS2A is 235 pc \citep{Hirota08}. Deuterium fractionation of formaldehyde and methanol in the envelope has been studied by \citet{parise06a}. The  
deuterium enrichments (HDCO/H$_2$CO\,$\sim$\,0.17 and CH$_2$DOH/CH$_3$OH\,$\sim$\,0.62) are higher than in IRAS 16293-2422 \citep{Parise04}. 

The H$_2$O line emission in IRAS2A was recently observed with the \textit{Herschel} telescope \citep{Kristensen10}.

\subsection{Observations} \label{observations}
\begin{table*}[!htbp]
\begin{center}
\renewcommand{\arraystretch}{0.9}
\caption{Parameters of the HDO observations}
\label{table1}
\begin{tabular}{lllllcllrrr}
\hline \hline
\noalign{\smallskip}
Telescope&Transition  & Frequency &  E$_{\mbox{\scriptsize{up}}}$ & Beam &  $B_{\mbox{\scriptsize{eff}}}$ &rms$^b$  &  Resolution &$T_{\mbox{\scriptsize{peak}}}^b$&$\Delta$v&$\int$$T_{\mbox{\scriptsize{mb}}}$dv\\
&& GHz & K & $''$& & mK &  km/s &mK & km/s & K km/s\\
\noalign{\smallskip}
\hline
\noalign{\smallskip}
\multicolumn{11}{c}{NGC1333-IRAS2A}\\
\noalign{\smallskip}
\hline
\noalign{\smallskip}
IRAM&1$_{1,0}$--1$_{1,1}$ & 80.578 & 46.8 		& 31.2&0.78 	& 11.6 	& 0.15	&17.8&3.9$\pm$1.1 &0.07$\pm$0.02\\
IRAM&3$_{1,2}$--2$_{2,1}$ & 225.897 & 167.7 	&11.1&0.57  	& 24.7 & 0.10	&113.5	&4.2$\pm$0.3&	0.50$\pm$0.03\\
IRAM&2$_{1,1}$--2$_{1,2}$ & 241.561 & 95.3 	&10.4&0.46 	& 30.8  &0.19	&98.6&4.1$\pm$0.5&0.43$\pm$0.05\\
JCMT&1$_{0,1}$--0$_{0,0}$  & 464.924 & 22.3 	& 10.8&0.44$^a$  	& 125 	&0.20&305.8&6.2$\pm$0.8&2.00$\pm$0.23\\
APEX&1$_{1,1}$--0$_{0,0}$  & 893.639 & 42.9 	& 7.0&0.35  	& 210  &0.25&...&...&...\\
\noalign{\smallskip}
\hline
\noalign{\smallskip}
\multicolumn{11}{c}{Outflow-position}\\
\noalign{\smallskip}
\hline
\noalign{\smallskip}
IRAM&1$_{1,0}$--1$_{1,1}$ & 80.578 & 46.8 		& 31.2&0.78 	& 11.5 	& 0.15&...&...&$\leq$0.03$^c$\\
IRAM&3$_{1,2}$--2$_{2,1}$ & 225.897 & 167.7 &11.1&0.57 		& 21.1 	&0.10&...&...&$\leq$0.04$^c$\\
JCMT&1$_{0,1}$--0$_{0,0}$  & 464.924 & 22.3 	& 10.8&0.44$^a$  	& 241 	&0.20&...&...&$\leq$0.72$^c$\\
\noalign{\smallskip}
\hline
\end{tabular}
\end{center}
$^a$Values as observed on Jupiter during the second semester of 2004 with the RxW-C receiver.\\
$^b$T$_{mb}$ scale\\
$^c$3$\sigma$ upper limit of integrated flux with assumption that $\Delta$v is 5 km/s.
\end{table*}

\begin{figure}[!htbp]
\includegraphics[width=9cm,angle=0]{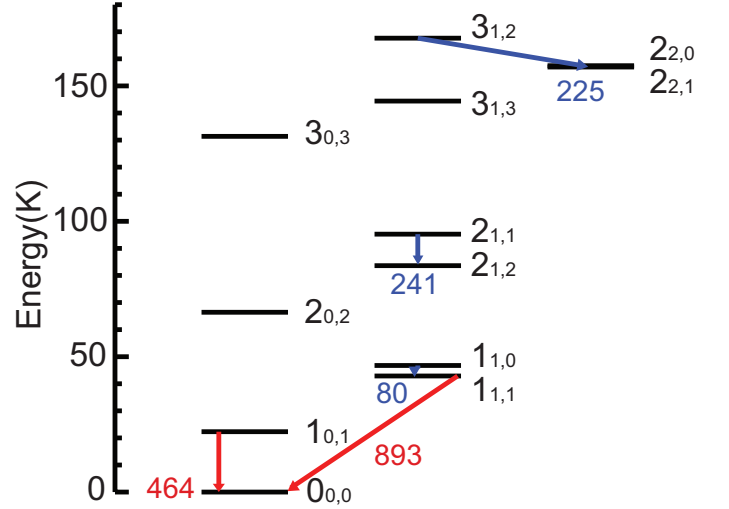}
\caption{HDO energy levels. The transitions that were observed in this work are marked with arrows (frequency in GHz).}
\label{fig1}
\end{figure}

Observations were carried out at three different telescopes. Using the IRAM 30m telescope,
we observed the 80, 225, and 241\,GHz lines (see Figure \ref{fig1}) toward IRAS2A, at position $\alpha_{2000}$=03$^h$28$^m$55$^s$.60
and $\delta_{2000}$=31$^o$14$^m$37$^s$.00, on 2004 Nov. 26 under PWV\,$\sim$\,2mm.
The focus was checked on Uranus, and the local pointing on 0333+321, leading to an
uncertainty in the pointing of less than 3$''$ (rms). We used the wobbler switching mode, with a throw of 90$''$.
Typical system temperatures varied in the range 115--150 K at 80\,GHz, 240--310\,K at 225\,GHz, and 375--525\,K at 241 GHz. During a second IRAM run, we observed an offset position relative to the source, to check for a possible contribution of the outflow. We targeted the 80GHz and 225GHz lines on 2005 Apr. 9, toward the position $\alpha_{2000}$=03$^h$29$^m$01$^s$.00 and $\delta_{2000}$=31$^o$14$^m$20$^s$.

The 464\,GHz observations were retrieved from the JCMT archive (project M04BN06). The 1$_{0,1}$-0$_{0,0}$ line was observed toward the position of IRAS2A as well as toward the outflow on 2004 Sep. 27, under an opacity $\tau_{\rm 225\,GHz}$=0.05.

The 893\,GHz line was observed using the CHAMP$^+$ multi-pixel receiver on APEX in Sep. 2008. The central 
pixel was centered on position $\alpha_{2000}$=03$^h$28$^m$55$^s$.40
and $\delta_{2000}$=31$^o$14$^m$35$^s$.00. Local pointing was checked in CO(6-5) on IK-Tau. We used the
wobbler symmetric switching mode, with a throw of 120$''$, resulting in OFF positions at 240$''$ from the source.
The total integration time is 95 minutes. The CO(6-5) line, observed in parallel on the 650\,GHz receiver, is clearly detected with a peak $T_{\mbox {\scriptsize A}}^*$ of order of 3\,K, and is presented in \citet{Kristensen10}.

Table \ref{table1} lists the characteristics of the HDO observations. The unit of flux here was converted from $T_{\mbox {\scriptsize A}}^*$ to $T_{\mbox {\scriptsize mb}}$ using the beam efficiencies indicated in the Table, which we took from the IRAM\footnote{http://www.iram.es/IRAMES/telescope/telescopeSummary/

telescope\_summary.html} and JCMT\footnote{http://www.jach.hawaii.edu/JCMT/spectral\_line/Standards/beameff.html} websites. The beam efficiency for CHAMP$^+$ was measured in Oct. 2007 on Mars.

\begin{figure}[!htbp]
\includegraphics[width=9cm,angle=0]{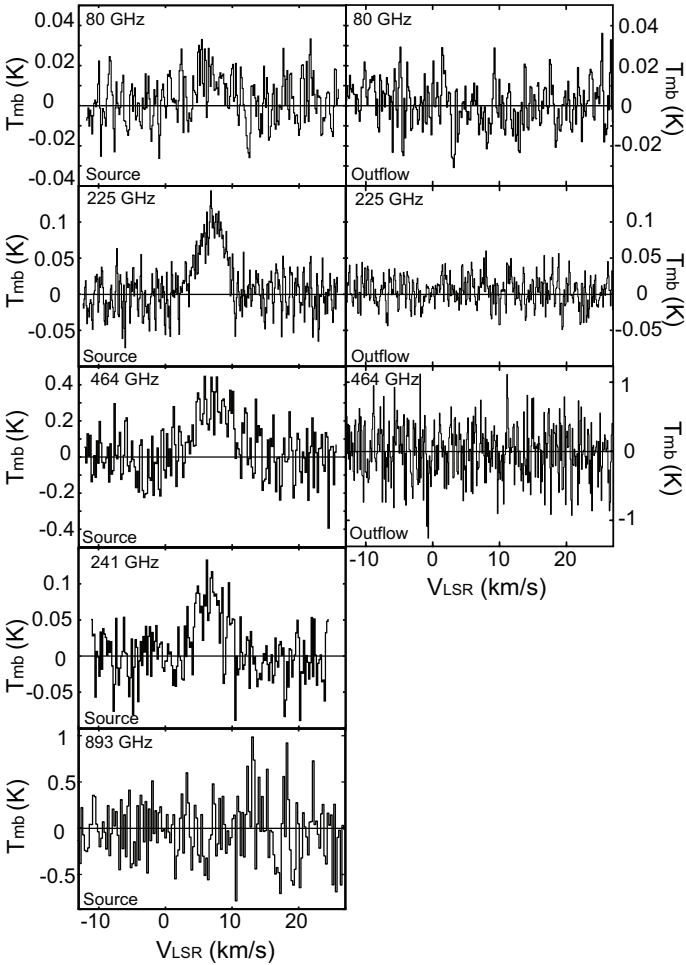}
\caption{Observed HDO spectra. The words in the lower left side of each spectrum indicate the observed position: source (on source) and outflow (offset-position).}
\label{fig2}
\end{figure}

\subsection{Results} \label{results}
The spectra observed toward IRAS2A and the offset outflow position are presented in Figure \ref{fig2}. The transitions from 80 GHz to 464 GHz are detected on-source, while we only have an upper-limit for the second ground transition, at 893 GHz. No line is detected at the offset position (see Table \ref{table1} and Figure \ref{fig2}). 

\section{Modeling}
\begin{figure}[!htbp]
\includegraphics[width=8 cm,angle=0]{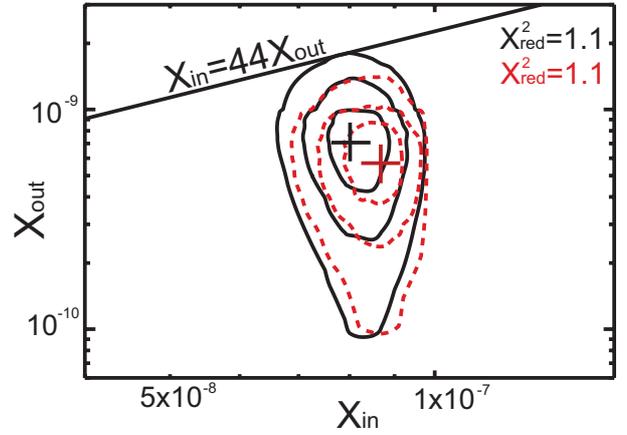}
\caption{x${^{\tiny\rm{HDO}}_{\rm in}}$ and x${^{\tiny\rm{HDO}}_{\rm out}}$
contours (1, 2 and 3$\sigma$) for the ${\rm \chi^2_{red}}$ for the different central star mass (M$_*$). The black-solid and red-dashed lines indicate the contours of M$_*=0.07$ M$_{\sun}$ and M$_*=0.15$ M$_{\sun}$. The symbols ``+'' correspond to the best-fit model of each case.} 
\label{fig3}
\end{figure}
Because no emission is detected at the targeted outflow position, we use an envelope model to fit the observational results, excluding any possible contribution of outflows or shocks. To analyze the HDO data, we used the 1D Monte Carlo code, RATRAN, developed by \citet{Hogerheijde00b}. The adopted density, opacity, and temperature profiles are derived with the 1D DUSTY code \citep{Ivezic97} following the procedure of \citet{Jorgensen02} \citep{Yildiz10}. In addition, we assumed the free-fall velocity field, $v=(\frac{2GM_*}{r})^{\frac{1}{2}}$ and adopted the somewhat arbitrary values for the central mass, M$_*=0.07$ and 0.15 M$_{\sun}$, consistent with the low mass inferred for the central object by \citet{Brinch09}. We will discuss this choice in Section 4.1. 
The emission was modeled in terms of a jump model, where the fractional abundances of deuterated water, relative to H$_2$, in the inner part of the source ($T\,>\,$100\,K, assumed evaporation temperature) and in the outer part ($T\,\le\,$100\,K) are two free parameters, x${^{\tiny\rm{HDO}}_{\rm in}}$ and x${^{\tiny\rm{HDO}}_{\rm out}}$, respectively. We performed an $\chi^2$ analysis for x${^{\rm \tiny{HDO}}_{\rm in}}$ ranging from 1$\times$10$^{-10}$ to 2$\times$10$^{-7}$ and for x${^{\tiny\rm{HDO}}_{\rm out}}$ ranging from 9$\times$10$^{-12}$ to 5$\times$10$^{-8}$.
Contrarily to the study by \citet{Parise05a}, who only modeled the integrated flux of HDO lines, we here intend to model as well the line velocity profiles. In order to get a higher signal-to-noise ratio and a consistent weighting of each spectra, we smoothed the observed spectra to similar resolutions between 0.4 -- 0.6 km/s. The $\chi^2$ was then computed on the profile of these smoothed spectra. Here the definition of $\chi^2$ is $\Sigma\frac{ (\mbox{\footnotesize T}_{\mbox{\scriptsize mb,ob}}-\mbox{\footnotesize T}_{\mbox{\scriptsize mb,mod}})^2}{\footnotesize \sigma^2}$, where the sum is over each channel in each spectrum. The parameters of the fits and their results are listed in Table \ref{table2}. The $\sigma$ within this $\chi^2$ analysis includes the statistical errors and uncertainties in flux calibration ($\sim$20\%), but does not include any uncertainty in the adopted collisional rate coefficients used in the excitation calculation \citep{Green89}.
The black-solid and red-dashed lines in Fig. \ref{fig3} present the contours delimitating the 1$\sigma$ (68.3\%), 2$\sigma$ (95.4\%) and 3$\sigma$ (99.7\%) confidence intervals. These contours are derived with the method described in \citet{Lampton76}, who shows that the $\Delta S=\rm \chi^2-\rm \chi^2_{min}$ random variable follows  a $\chi^2$ distribution with p variables, p being the number of parameters. Here p=2 (X$_{in}$ and X$_{out}$), so that these contours correspond to ${\rm \chi^2}$\,=\,${\rm \chi^2_{min}}$+2.3, ${\rm \chi^2_{min}}$+6.17 and ${\rm \chi^2_{min}}$+11.8. Here we use $^{``}$1$\sigma$$^{"}$, $^{``}$2$\sigma$$^{"}$, and $^{``}$3$\sigma$$^{"}$ in analogy with gaussian-distributed random variables.

\begin{table*}
\renewcommand{\arraystretch}{0.9}
\caption{Fitting parameters and their results}
\label{table}
\begin{tabular}{ccccccl}
\hline \hline
\noalign{\smallskip}
Central star mass$^a$&Doppler parameter $b^a$&\multicolumn{2}{c}{Best fit HDO abundances}&${\rm \chi^2_{red}}$&\multicolumn{2}{c}{HDO abundances confidence interval (3$\sigma$)}\\
\noalign{\smallskip}
\hline
\noalign{\smallskip}
(M$_{\sun}$)&(km/s)&inner region&outer region&& inner region&outer region \\
\noalign{\smallskip}
\hline
\noalign{\smallskip}
0.07 &2.0  &$8.0\times10^{-8}$  &$7.0\times10^{-10}$&1.1&$6.6\times10^{-8}$--$1.0\times10^{-7}$&$9\times10^{-11}$--$1.8\times10^{-9}$\\
0.15&1.0  &$8.6\times10^{-8}$  &$5.5\times10^{-10}$&1.1&$6.8\times10^{-8}$--$1.1\times10^{-7}$&$9\times10^{-11}$--$1.5\times10^{-9}$\\
\noalign{\smallskip}
\hline
\end{tabular}\\
$^a$ Fixed parameters
\label{table2}
\end{table*}

\begin{figure}[!htbp]
\includegraphics[width=9cm,angle=0]{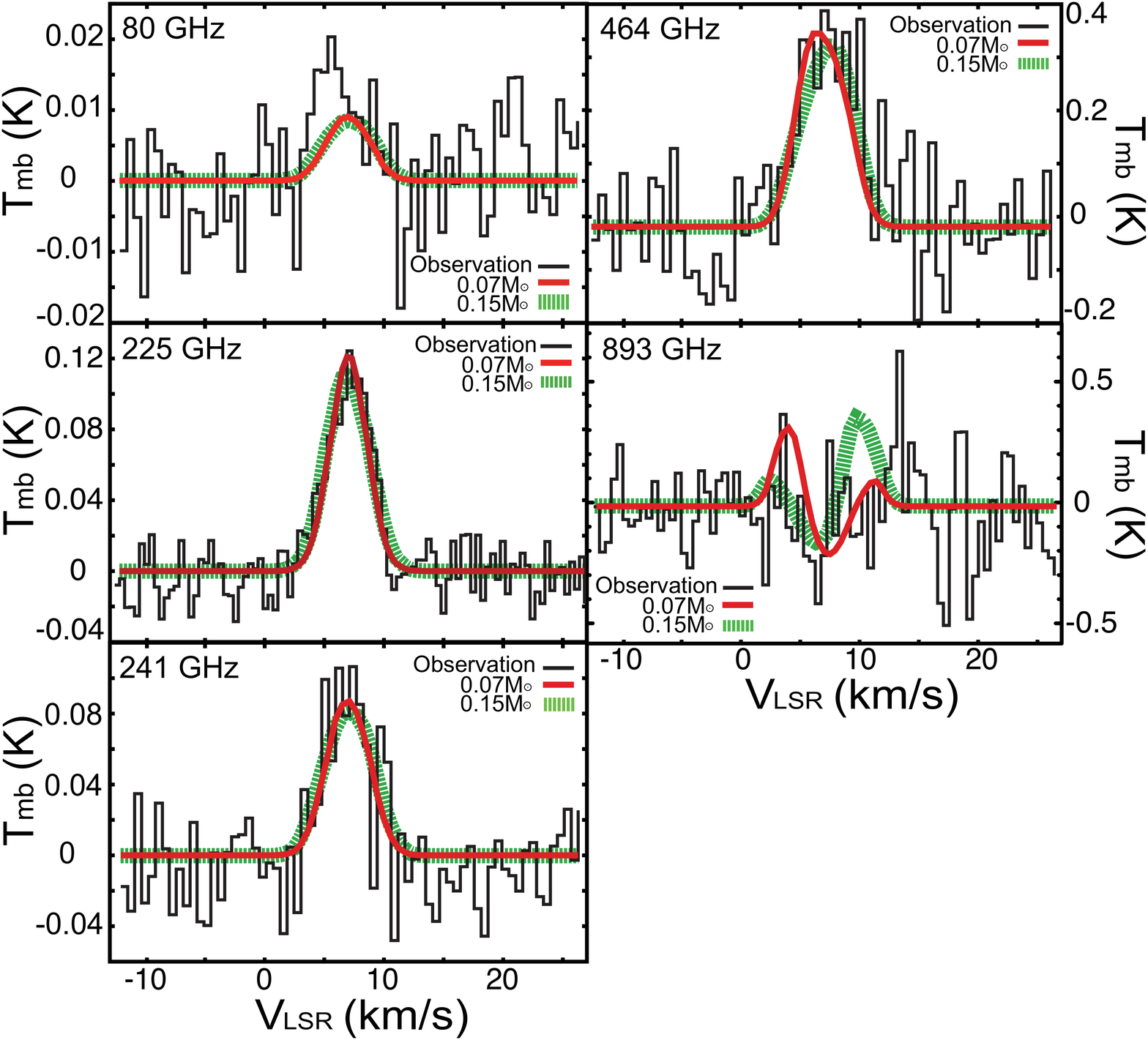}
\caption{Comparison of the observed spectra smoothed to 0.4 -- 0.6 km/s spectral resolution with the results of our best-fit models for two different masses of the central protostar. The observed line emission is shown in black lines in each spectrum. The red and green lines show the results of the models with the assumptions that M$_*$ is 0.07 M$_{\sun}$ and 0.15 M$_{\sun}$, respectively.}
\label{fig4}
\end{figure}

The obtained minimum ${\rm \chi^2_{red}}$ is not far from 1, which indicates that the model assuming a simple collapsing envelope is a reasonable description of the data. The inner and outer fractional abundance is well constrained by our data at the 3$\sigma$ level. The smoothed spectra are overlaid with the modeling results of the best fit with two different central-star masses in Fig. \ref{fig4}.

\begin{table*}[!htbp]
\renewcommand{\arraystretch}{1.2}
\caption{Comparison of HDO fractional abundance between NGC1333-IRAS2A and IRAS 16293-2422}
\label{table}
\begin{center}
\begin{tabular}{l|cc|rc|cccc}
\hline \hline
\noalign{\smallskip}
Source&\multicolumn{2}{c}{best fit}&\multicolumn{2}{c}{confidence interval (3$\sigma$)}&\multicolumn{2}{c}{$\frac{\mbox{HDO}}{\mbox{H$_2$O}}$}&$\frac{\mbox{HDCO}}{\mbox{H$_2$CO}}$$^b$&$\frac{\mbox{CH$_2$DOH}}{\mbox{CH$_3$OH}}$$^b$\\
\noalign{\smallskip}
\hline
\noalign{\smallskip}
&x${^{\tiny\rm{HDO}}_{\rm in}}$&x${^{\tiny\rm{HDO}}_{\rm out}}$  & x${^{\tiny\rm{HDO}}_{\rm in}}$&x${^{\tiny\rm{HDO}}_{\rm out}}$&Inner &outer $(3\sigma)$&---&---\\
\noalign{\smallskip}
\hline
\noalign{\smallskip}
IRAS2A$^a$ &$8\times10^{-8}$&$7\times10^{-10}$  &$6.6\times10^{-8}$--$1.0\times10^{-7}$ &$9\times10^{-11}$--$1.8\times10^{-9}$&$\geq$ 0.01$^c$& 0.07$^{+0.11c}_{-0.06}$ &0.17$^{+0.12}_{-0.08}$&0.62$^{+0.71}_{-0.33}$\\
IRAS 16293$^d$&$1\times10^{-7}$  &$1.5\times10^{-10}$  &$7\times10^{-8}$--$1.3\times10^{-7}$  &$\leq 1.0\times10^{-9}$&0.03&$\leq 0.002$ &0.15$\pm$0.07&0.37$^{+0.38}_{-0.19}$\\
\noalign{\smallskip}
\hline
\noalign{\smallskip}
Orion KL&...  &...  &... &...&\multicolumn{2}{c}{0.02$^e$}&0.14$^f$&0.04$^g$\\
\noalign{\smallskip}
\hline
\end{tabular}
\end{center}
 $^a$The HDO fractional abundance was derived with M=0.07 M$_{\sun}$.\\
$^b$\citet{parise06a}\\
$^c$The fractional abundances of water taken here are x${^{\tiny\rm{H_2O}}_{\rm in}}\leq$$1\times10^{-5}$ and x${^{\tiny\rm{H_2O}}_{\rm out}}=$$1\times10^{-8}$, as constrained by \citeauthor{Visser} et al. in prep. from the $1_{11}-0_{00}$, $2{02}-1{11}$, and $2_{11}-2_{02}$ H$_2^{16}$O lines and the upper-limit on the $1_{11}-0_{00}$ H$_2^{18}$O line presented in Kristensen et al. (2010).\\
$^d$\citet{Parise05a}\\
$^e$\citet{Bergin10}\\
$^f$\citet{Turner90}\\
$^g$\citet{Jacq93}\\
\label{table3}
\end{table*}

\section{Discussion}

The first results of our model are that the observations are well described by the adopted envelope model, and that the HDO abundance profile displays a jump by two orders of magnitude between the cold and the warm envelope for the best fit. Below we first discuss the choice of the velocity field and the Doppler $b$-parameter, which is the turbulence broadening.

\subsection{The high turbulence broadening}
Although the minimum ${\rm \chi^2_{red}}$ value shows that the simplest model (i.e. an infalling envelope) describes our data well, the required turbulence broadening b of 2 km/s may seem unrealistically high for the best model (M=0.07 M$_{\sun}$). We note however that this $b$ parameter is degenerate within the noise of our observations with the choice of the free-fall velocity profile in the envelope. Indeed, increasing the central mass (which is anyway mainly unconstrained) to 0.15 M$_{\sun}$ allows us to reproduce the linewidths with a significantly lower $b=1$ km/s. The minimum ${\rm \chi^2_{red}}$ in this case is the same as the model with central mass 0.07 M$_{\sun}$, and the x${^{\rm \tiny{HDO}}_{\rm in}}$ and x${^{\rm \tiny{HDO}}_{\rm out}}$ of the best fit model are $8.6\times10^{-8}$ and $5.5\times10^{-10}$, which are still within the 1$\sigma$ of previous results (Fig. \ref{fig3}). This shows that the constraints on x${^{\rm \tiny{HDO}}_{\rm in}}$ and x${^{\rm \tiny{HDO}}_{\rm out}}$ do not depend much on the chosen velocity profile, and that the turbulence is barely constrained in the models.

\subsection{The fractional abundance of HDO}
 The low value of ${\rm \chi^2_{red}}$ = 1.1 for the best-fit model shows that the assumption of a jump profile for the abundance in a spherical envelope is appropriate to fit our data. The abundance of HDO jumps by at least a factor of 44 at the 3$\sigma$ confidence level (see Fig. \ref{fig3}). Ices are therefore shown to evaporate from the grains in the inner part of the envelope in IRAS2A. This property has also been shown on HDO and H$_2$O in IRAS 16293-2422 \citep{Parise05a,Ceccarelli00a}. 
Moreover, the best-fit HDO abundances of the two cases are also similar to the values of HDO fractional abundances in IRAS 16293-2422 (see Table \ref{table3}). 
This suggests that the mechanisms to form HDO could be the same in both these solar-type class 0 protostars, even if the two sources are located in different star-forming regions (Perseus v.s. Ophiuchus), and that the case of IRAS 16293-2422 is not peculiar. The present study provides the second detailed multi-transition HDO analysis toward a low-mass protostar, and the study of a greater number of sources would be essential to generalize this result.

\begin{figure}[!htbp]
\includegraphics[width=8.8cm,angle=0]{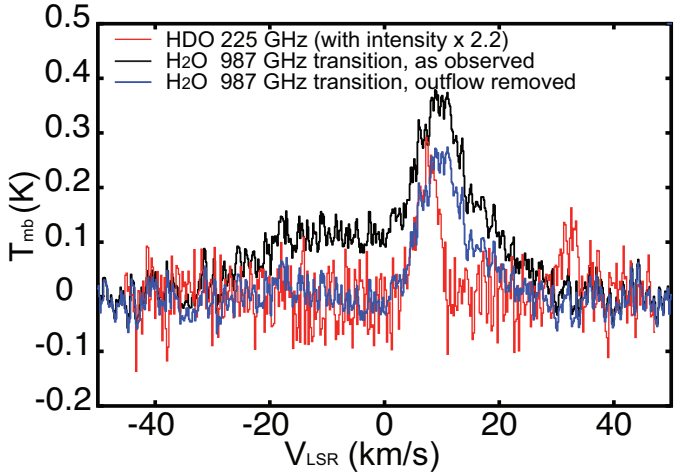}
\caption{Comparison of the HDO emission with H$_2$O emission with/without wing. The red line indicates the emission line of the HDO transition at 225 GHz . The black and blue line show the line emission of the H$_2$O transition at 987 GHz with and without outflow component. The H$_2$O data are taken from \citet{Kristensen10}.}
\label{fig6}
\end{figure}

\subsection{Water deuterium fractionation in low-mass protostars}

In order to derive the HDO/H$_2$O ratio in the IRAS2A envelope, we need to infer the H$_2$O abundance profile in the quiescent envelope.
Figure \ref{fig6} compares the spectra of the 225 HDO line and the 987 H$_2$O line (Kristensen et al. 2010). Both the total water emission and the emission where the broad outflow Gaussian component was removed are shown. In this figure we multiplied the intensity of 225 HDO line with a factor of 2.2 to compare it with the H$_2$O emission.
Obviously the H$_2$O line is broader than the HDO line (FWHM\,=\,9.5 km s$^{-1}$ versus up to 6 km s$^{-1}$ for HDO), even where the broad outflow component was removed. This medium-broad component has been interpreted to come from small-scale
shocks because of the interaction of the jet and wind with the dense inner envelope on scales of $\sim 1''$ \citep{Kristensen10}. In addition, the H$_2$O emission from the warm inner envelope was suggested to be optically thick \citep{Kristensen10}. Therefore, the H$_2$O lines do not probe the quiescent envelope and HDO lines may not probe the same gas.

To obtain upper limits on the quiescent H$_2$O abundance, Visser et
al. in prep. used deep integrations on various H$_2^{18}$O lines presented by
Kristensen et al. (2010). The non-detection of narrow H$_2^{18}$O features limits
the inner H$_2$O abundance to $\lesssim 10^{-5}$, whereas the presence of
narrow deep H$_2$O self-absorption gives an outer H$_2$O abundance of $\sim
10^{-8}$. 

The values of the HDO/H$_2$O ratio within the present constraints are listed in Table \ref{table3}. The HDO/H$_2$O ratios in the inner warm and the outer cold envelope are found to be more than 1\% and are in the range of 0.9\%--18\% (3$\sigma$). The D/H ratio of water in the outer envelope is roughly one order of magnitude lower than the ratio of methanol, which confirms the trend measured in IRAS 16293$-$2422. Obviously, better constraints on the H$_2$O abundances in the inner warm envelope are required to see if the deuterium enrichment in water is significantly lower than in methanol. On the other hand, the similar HDO/H$_2$O ratio and the one order magnitude higher CH$_2$DOH/CH$_3$OH ratio of IRAS2A and Orion KL suggest that the environment plays a role in the chemistry.

The HDO/H$_2$O ratio in the inner envelope is found to be more than one order of magnitude higher than the ratio measured in comets, as for IRAS 16293$-$2422. If this high deuterium enrichment of water is typical of low-mass protostellar envelopes, and if the HDO/H$_2$O ratio measured in the cometary coma is representative of the cometary ice composition, then this would imply that the deuterium fractionation of water is reprocessed at some point between the protostellar envelope and the cometary ice. Reprocessing of the gas in the hot corino warm environment is expected to reduce the deuterium fractionation on typical timescales of 10$^5$ yrs in environments of density $\sim$10$^6$ cm$^{-3}$ \citep{Charnley97}. However, in this case, re-condensation of the gas on the grains would be required. Isotopologue exchanges are also possible at the surface of the grains \citep[see e.g. ][]{Ratajczak09}, without requiring evaporation. This may however not be enough to deplete the deuterium level in water, because water is the main component of water ices. A mechanism to get rid of deuterons from the grain surface would need to be invoked. Because of the high abundance of water, this mechanism should involve efficient exchange reactions. A suggestion would be to study in the laboratory the possibility of  deuteron exchange between HDO and H$_2$ at the grain surfaces. 

\begin{figure}[!htbp]
\includegraphics[width=8.8cm,angle=0]{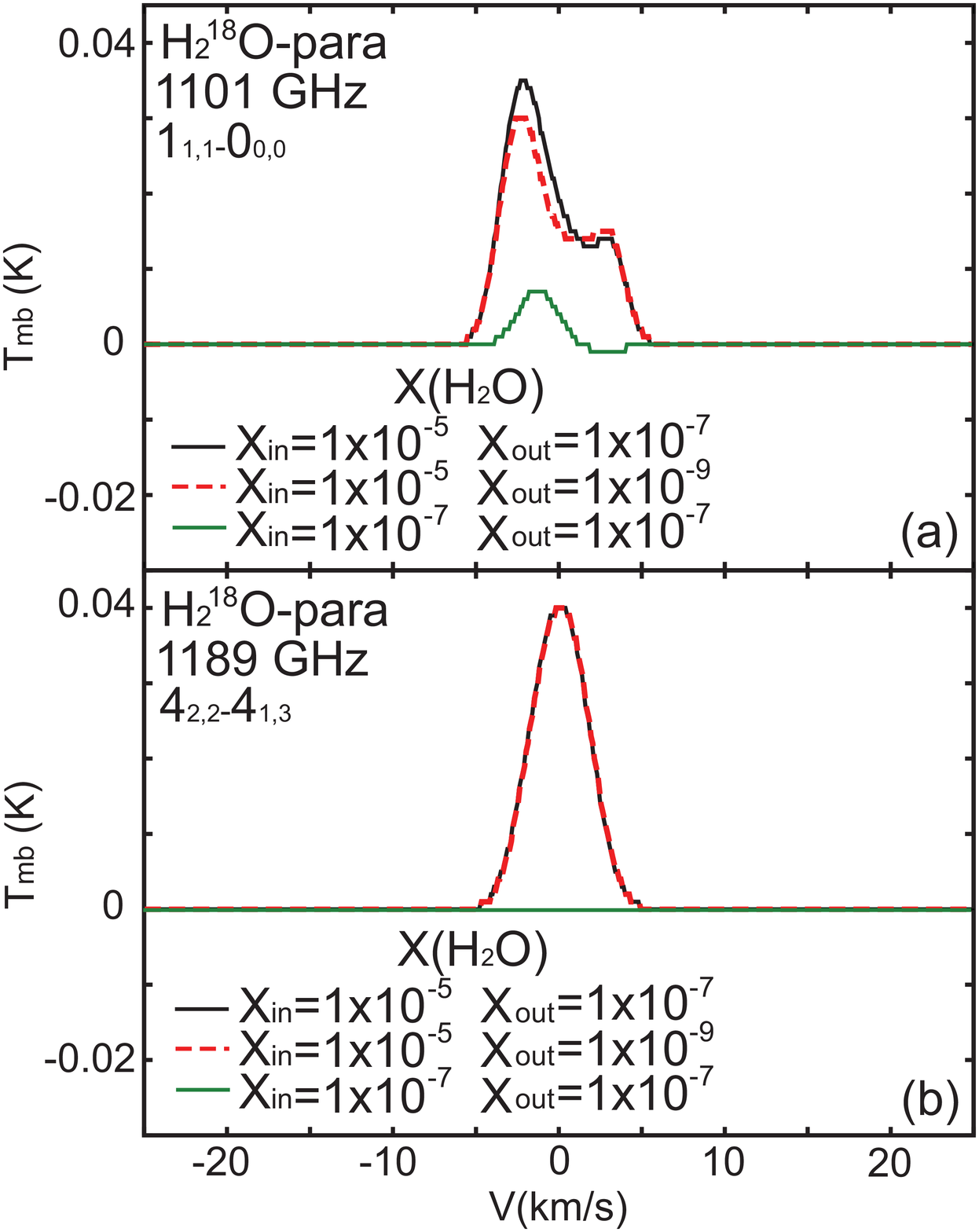}
\caption{Model predictions for the chosen H$_2^{18}$O lines. X$_{in}$ and X$_{out}$ represent the water (H$_2$O) abundance in the inner warm ($>$ 100 K) and outer cold ($\,\le\,$ 100 K) envelope. A standard ortho/para ratio of three was assumed.}
\label{fig7}
\end{figure}

Improving the observational constraints on the HDO/H$_2$O ratio in protostellar envelopes is essential to allow us to understand the implications in terms of chemical evolution in the protosolar nebula. 
Figure \ref{fig7} shows the RATRAN modeling prediction for two H$_2^{18}$O lines, for a set of reasonable water abundances. We made the assumption that the standard ortho/para ratio is 3 and the $^{18}$O/$^{16}$O ratio is 2.05$\times$10$^{-3}$. Figure \ref{fig7}b shows that the H$_2^{18}$O 1189 GHz line emission is much more sensitive to the inner fractional abundance of water than the H$_2^{18}$O 1101 GHz line. This line is therefore a good candidate to target with the \textit{Herschel} telescope to further constrain the inner water abundance.

\section{Conclusion}
We presented the observation of five HDO lines toward the solar-type class 0 protostar NGC1333-IRAS2A using the IRAM 30m, JCMT and APEX telescope. Four of them are clearly detected, while the second ground-state transition line at 893 GHz is not. Because there are no detections at the outflow position at 80, 225, and 464 GHz, we assumed for the modeling that most emission comes from the envelope. 

We modeled the emission with the RATRAN radiative transfer code, assuming a step profile for the HDO abundance.  We derived the HDO abundance in the inner and outer parts of the envelope to be x${^{\rm \tiny{HDO}}_{\rm in}}=8\times10^{-8}$ and  x${^{\tiny\rm{HDO}}_{\rm out}}=7\times10^{-10}$. This result shows that the HDO abundance is enhanced in the inner envelope because of the ices' evaporation from the grains, as for IRAS 16293-2422. Moreover, the values are similar to those in IRAS 16293-2422, which suggests that the pathway of the formation of HDO is the same in low-mass class 0 protostars. The study of a larger sample of sources is needed to generalize this result.

The HDO/H$_2$O abundance ratio is found to be $>$1\% in the inner envelope. A better constraint on the H$_2$O abundance in the inner envelope is required to derive the deuterium enrichment in the warm envelope.

\begin{acknowledgements}
The authors warmly thank the anonymous referee for a very constructive report, which greatly improved the treatment of the confidence intervals for the HDO inner and outer abundances. The authors are grateful to the WISH team for providing access to the H$_2$O data. F.C. Liu and B. Parise are funded by the German \textit{Deutsche Forschungsgemeinschaft, DFG} Emmy Noether project number PA1692/1-1. Astrochemistry in Leiden is supported by the Netherlands Research
School for Astronomy (NOVA) and a Spinoza grant from the Netherlands
Organization for Scientific Research (NWO).
\end{acknowledgements}

\bibliographystyle{aa}
\bibliography{15519IRAS2a.bib}

\end{document}